# How time window influences biometrics performance: an EEG-based fingerprints connectivity study

Luca Didaci[1*], Sara Maria Pani[2], Claudio Frongia[1] and Matteo Fraschini[1]

[1] Department of Electrical and Electronic Engineering, University of Cagliari, via Marengo 2, 09123 Cagliari, Italy
[2] Department of Medical Science and Public Health, University of Cagliari, 09123 Cagliari, Italy

[*] Corresponding author

**Corresponding author**:
Prof. Luca Didaci
Department of Electrical and Electronic Engineering
University of Cagliari
via Marengo 2
09123 Cagliari
Italy
e-mail: didaci@unica.it
telephone: +39 070 675 5844

**Abstract**

EEG-based biometric represents a relatively recent research field that aims to recognize individuals based on their recorded brain activity by means of electroencephalography (EEG). Among the numerous features that have been proposed, connectivity-based approaches represent one of the more promising methods tested so far. In this paper, we investigate how the performance of an EEG biometric system varies with respect to different time windows to understand if it is possible to define the optimal duration of EEG signal that can be used to extract those distinctive features. Overall, the results have shown a pronounced effect of the time window on the biometric performance measured in terms of EER (equal error rate) and AUC (area under the curve), with an evident increase of the biometric performance with an increase of the time window. In conclusion, we want to highlight that EEG connectivity has the potential to represent an optimal candidate as EEG fingerprint and that, in this context, it is very important to define a sufficient time window able to collect the subject specific features. Moreover, our preliminary results show that extending the window size beyond a certain maximum does not improve biometric systems' performance.

**Keywords**: EEG, biometric, connectivity, time window

## Introduction

EEG-based biometric represents a relatively recent research field that aims to recognize individuals based on their recorded brain activity by means of the electroencephalography (EEG), and it is related to the more specific topic of the evaluation of individual fingerprint of a human functional connectome (Amico and Goñi, 2018). During the last few years, there has been growing evidence that EEG signals, which reflect neural responses to various kinds of stimuli, including resting-state activity, can be successfully used to derive distinctive features to develop high-performance, secure, and robust biometric recognition systems (Fidas and Lyras, 2023; Jalaly Bidgoly et al., 2020; Saia et al., 2023; Zhang et al., 2021).

Among the numerous features proposed to investigate how the EEG signals behave in terms of brain activity fingerprint, connectivity-based approaches represent one of the more promising methods tested so far (Fraschini et al., 2019, 2015; Wang et al., 2020). In particular, phase-based connectivity methods, which allow the estimation of the correlation between different EEG channels (or the corresponding reconstructed sources) based on the phase synchronization of EEG signals, outperformed other connectivity and power-based approaches (Fraschini et al., 2019; Pani et al., 2020), thus representing a valid solution to develop high-performance EEG biometric systems. In this context, the PLI (phase lag index) (Stam et al., 2007) and the PLV (phase locking value) (Lachaux et al., 1999) connectivity methods have shown to be good candidates as EEG-based features to recognize individuals in several experimental scenarios.

Despite their wide application and inherent advantages over more common biometrics, EEG-based biometric systems still have some serious drawbacks, including, but not limited to, the need of standardized pipelines for the analysis (Jalaly Bidgoly et al., 2020). In particular, it has been shown that the definition of the EEG time-window represents a crucial issue in the EEG analysis. Its length may strongly affect several EEG features, especially those based on connectivity methods (Basti et al., 2022; Fraschini et al., 2016). On the other hand, understanding the optimal time-window in an EEG biometric system is of huge interest because it allows one to define, at least for brain signals, the minimum required amount of time necessary to have the system properly working with satisfactory performance.

In this paper, we investigate how the performance of an EEG biometric system, based on two common and widely used phase connectivity methods, varies with respect to different time-windows with the aim to understand if, at least for EEG fingerprints approaches, it is possible to define the optimal duration of EEG signal that can be used to extract those distinctive features. We evaluated our hypothesis using a public EEG dataset (Goldberger Ary L. et al., 2000) consisting of recordings from 109 subjects recorded with a high-density EEG system that can be downloaded at the following link: (http://physionet.org/pn4/eegmmidb/). The analysis was performed using both eyes-closed and eyes-open resting-state conditions, which have been previously shown to be able to develop high-performance EEG biometric systems (Fraschini et al., 2015). Finally, we also discuss the implications and limitations of this analysis, suggesting directions for future work related to this challenging research field.

## Materials and methods

The dataset.

In this study, we used a high-density EEG dataset (64 channels) that is publicly and freely accessible and includes 109 healthy subjects. The raw data are available on the PhysioNet website (http://physionet.org/pn4/eegmmidb/). This dataset has been previously employed for brain-computer interface and biometric applications. The EEG signals were acquired with a sampling rate of 160 Hz, referenced to the average of the ear-lobe electrodes and organized into several runs involving resting state, motor movement and imaginary tasks.

For the following analysis, we selected two resting-state runs (eyes open and eyes closed), each lasting 1 minute.

Pre-processing.

The raw EEG signals were band-pass filtered (without phase distortion (Delorme and Makeig, 2004)) using high beta (20–30 Hz) and gamma (30–45 Hz) frequency bands since other frequency content did not show relevant results in the context of EEG fingerprinting (Fraschini et al., 2019). Successively, for each condition (eyes open and eyes closed), each subject and each frequency band, the following steps were performed: (i) epoch segmentation; (ii) functional connectivity analysis, using both PLI and PLV methods; and (iii) performance evaluation.

Epoch segmentation.

The preprocessed signals were organized in epochs of different lengths, ranging from 0.5 s to 12 s using a step of 0.5 s, thus obtaining 24 different time window sizes. All the subsequent analysis was performed for these different time windows separately.

Functional connectivity analysis.

This step was performed by evaluating pair-wise statistical interdependence between EEG signals using the PLI (phase lag index) (Stam et al., 2007) and the PLV (phase locking value) (Lachaux et al., 1999) methods. The PLI is an index of asymmetry of the distribution of instantaneous phase differences between EEG channels. Its values range from 0 (no interaction, or interaction with zero phase lag) to 1 (maximum interaction), and it has been shown to be robust to signal spreads, linear mixing, and active reference. The PLV allows to detect transient phase locking values, independent of the signal amplitude and represents the absolute value of the mean phase difference between the two EEG channels.

Performance evaluation.

In a hypothetical scenario of user authentication, we used a feature vector containing the connectivity values (for PLI and PLV separately), extracted using the upper triangular part of the correlation matrix to compute the biometric performance.

As previously reported (Crobe et al., 2016; Fraschini et al., 2015), pair-wise similarity scores between feature vectors were estimated as $1/(1 + d)$, where $d$ represents the Euclidean distance. Genuine and impostor score distributions were finally used to compute the equal error rate (EER) and the area under the ROC curve (AUC), which allowed us to assess the performance for each band, each condition, and each time window.

Results

As summarized in Figure 1 and Figure 2, the results of the study clearly show an evident increase in the biometric performance, as expressed in terms of EER, with an increase of the time window, for all the experimental conditions (eyes-closed and eyes-open resting state), for both beta and gamma frequency bands and, even though with different magnitude, for both PLV and PLI methods. In particular, the best performance was obtained for the PLV approach in the gamma band for the eyes-open resting state condition with a minimum time window equal to at least 10.5 s (EER = 0.018) and for PLV approach in the beta band for the eyes-closed resting state condition again with a minimum time window equals to at least 10.5 s (EER = 0.035). In more detail, Figure 1 shows the effect of using different time windows, with a step of 0.5 s, on the EER for both PLV and PLI, for beta and gamma frequency bands and for the two resting state conditions.

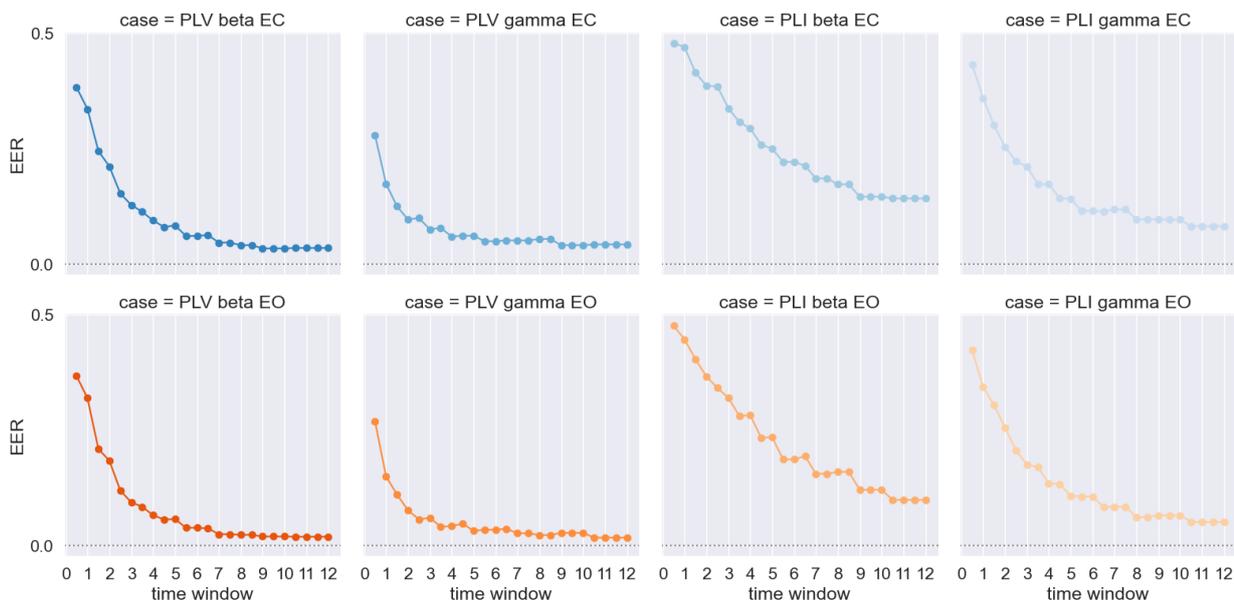

Figure 1. The effect of using different time windows, from 0.5 s to 12 s, with a step of 0.5 s, on the EER for both PLV and PLI, for beta and gamma frequency bands and for the two resting state conditions, namely eyes-closed (EC) and eyes-open (EO).

Moreover, Figure 2 shows the effect of using different time windows, always with a step of 0.5 s, on the AUC, again for both PLV and PLI, for beta and gamma frequency bands and for the two resting state conditions.

Finally, to better represent the effect of the time window on the computed performance, Figure 3 shows the score distributions for the two more extreme windows, the shorter of 0.5 s and the larger of 12 s, for the PLV methods, in the gamma frequency band for the eyes-open resting state condition. This result clearly shows how the time window affects the overall performance in terms of total overlap between genuine and impostor score distributions.

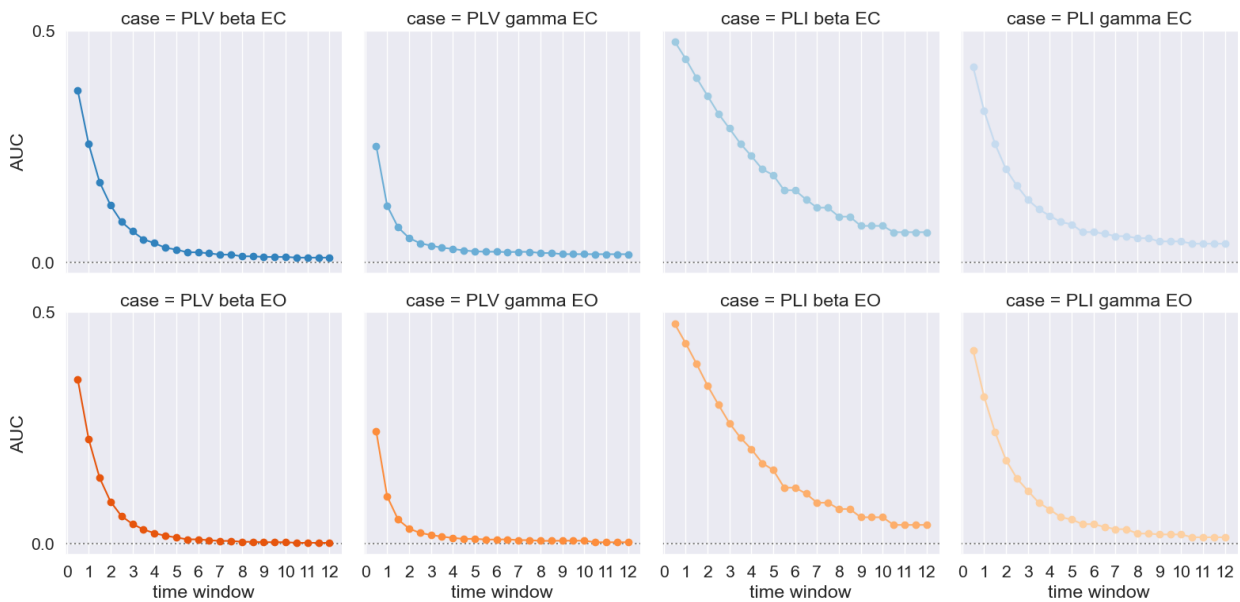

Figure 2. The effect of using different time windows, from 0.5 s to 12 s, with a step of 0.5 s, on the AUC for both PLV and PLI, for beta and gamma frequency bands and for the two resting state conditions, namely eyes-closed (EC) and eyes-open (EO).

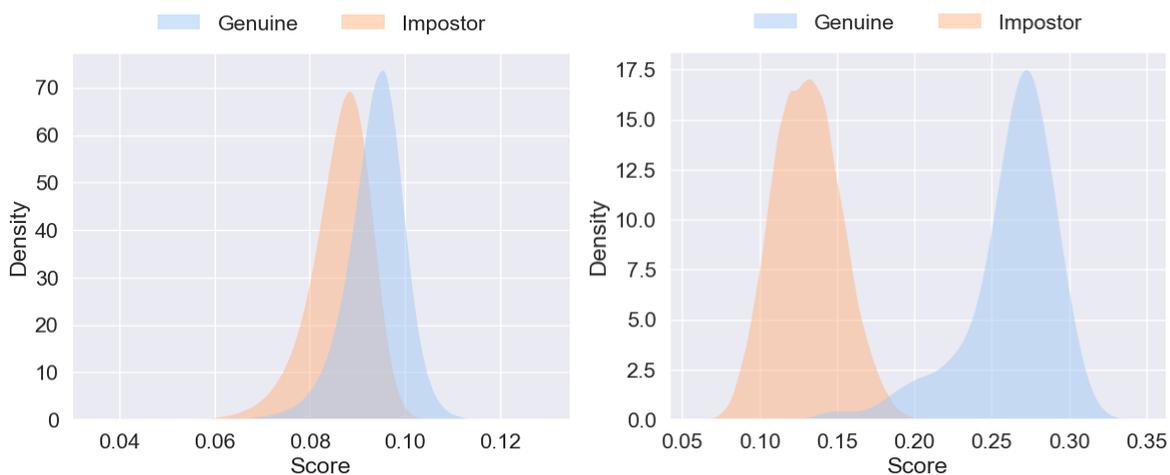

Figure 3. The score distributions for the two more extreme windows, the shorter of 0.5 s and the larger of 12 s, for the PLV methods, in the gamma frequency band for the eyes-open resting state condition.

**Discussion**

In the present study, we investigated the effect of the time window on a hypothetical EEG-based biometric system for user authentication. To test this research question, we used two well known, widely used connectivity metrics that have been shown to have good aptitude as subject specific fingerprints in resting state EEG (Fraschini et al., 2019; Pani et al., 2020). The analysis was performed on a freely available EEG dataset (Goldberger Ary L. et al., 2000) to easily reproduce the whole study.

Overall, as hypothesized, the reported results showed a pronounced effect of the time window on the biometric performance measured in terms of EER and AUC. In particular, we have observed that the EER varies from 0.478 for the smaller time window to 0.018 for the larger. The same result was observed for the AUC, which, again, varies from 0.476 for the smaller window to 0.002 for the larger one. These findings suggest that it is possible to obtain very high performance even with relatively short EEG time windows in the order of 10 seconds. This should allow the development of biometric systems that make use of very short EEG recordings with the consequence of limiting the quantity of data to be acquired.

In our opinion, these results align with previous studies (Fraschini et al., 2016) that have reported the effect of epoch length on connectivity metrics. In brief, we speculate that the poor performance obtained with short time windows mainly depends on the difficulty of correctly estimating the connectivity metrics from very small EEG epochs (shorter than 6-8 s).

In conclusion, we want to highlight that, to develop any EEG based biometric system, EEG connectivity represents an optimal candidate as EEG fingerprint and that, in this context, it is very important to define a sufficient time window able to collect the subject-specific features. Moreover, our preliminary results show that extending the window size beyond a certain maximum does not improve biometric systems' performance.

**Acknowledgements**

Matteo Fraschini, Claudio Frongia and Luca Didaci were in part funded by Fondazione di Sardegna under the project ``TrustML: Towards Machine Learning that Humans Can Trust'', CUP: F73C22001320007.